\documentclass[aps,prl,twocolumn,showpacs,superscriptaddress]{revtex4-1}
\usepackage{graphicx,amsmath,amssymb}
\usepackage[usenames]{color}
\begin{document}

\title{Quantum phase transition and spontaneous symmetry breaking in a nonlinear quantum Rabi model}

\author{Zu-Jian Ying }
\email{yingzj@lzu.edu.cn}
\affiliation{School of Physical Science and Technology $\&$ Key Laboratory for Magnetism and Magnetic Materials of the
Ministry of Education, Lanzhou University, Lanzhou 730000, China}
\affiliation{CNR-SPIN and Dipartimento di Fisica ``E. R. Caianiello'', Universit\`a di Salerno, 84084 Fisciano, Italy}

\author{Lei Cong}
\email{congl09@lzu.edu.cn}
\affiliation{School of Physical Science and Technology $\&$ Key Laboratory for Magnetism and Magnetic Materials of the
Ministry of Education, Lanzhou University, Lanzhou 730000, China}

\author{Xi-Mei Sun}
\affiliation{School of Physical Science and Technology $\&$ Key Laboratory for Magnetism and Magnetic Materials of the
Ministry of Education, Lanzhou University, Lanzhou 730000, China}

\begin{abstract}
The experimental advance on light-matter interaction into strong couplings has invalidated Jaynes-Cummings model and brought quantum Rabi model (QRM) to more relevance. The QRM only involves linear coupling via a {\it single}-photon process (SPP), while nonlinear {\it two}-photon process (TPP) is weaker and conventionally neglected. However, we find a contrary trend that enhancing the linear coupling might not suppress more the nonlinear effect but backfire to trigger some collapse of linear characters. Indeed, in strong SPP couplings a tiny strength of TPP may dramatically change properties of the system, like a symmetry spontaneous breaking. By extracting the ground-state phase diagram including both SPP and TPP, we find TPP in low frequency limit induces a quantum phase transition with continuity-discontinuity double faces, which split into two distinct transitions at finite frequencies and yields a triple point. Our analysis unveils a subtle SPP-TPP entanglement.
\end{abstract}
\pacs{ }
\maketitle


\textit{Introduction.--}Recently significant efforts in experiments have
pushed the exploration of fundamental quantum physics in light-matter
coupling systems toward the (ultra-)strong coupling regime \cite%
{Wallraff2004, Gunter2009, Niemczyk2010, Peropadre2010,FornDiaz2017,
Forn-Diaz2010, Cristofolini2012, Scalari2012,
Xiang2013,Yoshihara2017,Andersen2017,Kockum2017}. This experimental
enhancement on coupling strength has rendered the Jaynes-Cummings model less
valid\cite{Niemczyk2010, Forn-Diaz2010} and brought the quantum Rabi model
(QRM) \cite{rabi1936} more to the front of investigation for light-matter
interaction. Also theoretically the remarkable finding of integrability of
the QRM \cite{Braak2011} has added a great fuel to heat up the interest \cite%
{Solano2011,exact_chenqh,prx_xie,Batchelor2015,Ying2015,LiuM2017PRL,Hwang2015PRL,AnistropicShen2017,Ashhab2010,Ref-gc0,CongLei2017,Liu2015,ChenGang2012,FengMang2013}
in the model. The QRM is also particular for few-body quantum phase
transition (QPT) \cite%
{LiuM2017PRL,Ying2015,Hwang2015PRL,AnistropicShen2017,Ashhab2010,Ref-gc0,CongLei2017}%
, which can be bridged to QPTs in the thermodynamic limit\cite{LiuM2017PRL}.

Conventionally the QRM is a linear model which involves the coupling of a
qubit or spin-half system and a bosonic mode via a \textit{single}-photon
process (SPP) of absorption and emission, while the nonlinear \textit{two}%
-photon process (TPP) \cite{Prokhorov1965-TwoPhotonProcess}, is usually much
weaker and not taken into account. Nowadays, the nonlinear process has been
realized in different systems, e.g. Rydberg atoms \cite%
{Bertet2002-TwoPhotonProcess, Brune1987-TwoPhotonProcess} and quantum dots
\cite%
{Stufler2006-TwoPhotonProcess,Valle2010-TwoPhotonProcess,Verma2016-TwoPhotonProcess}
in microwave cavities, and enhanced in trapped ions \cite%
{Felicetti2015-TwoPhotonProcess,Puebla2017-TwoPhotonProcess} and
superconducting circuits\cite%
{Felicetti2018-mixed-TPP-SPP,Pedernales-PRL-2018,Bertet-Nonlinear-Experim-PRL-2005,Bertet-Nonlinear-Experim-Model-2005}%
. In such a situation, the issue of TPP-SPP competition arises. The
traditional way to see the TPP effect is to suppress the SPP\cite%
{Felicetti2018-mixed-TPP-SPP,Lange1996-SuppressSPP,Ota2011-SuppressSPP},
while knowledge is lacking in wondering reversely the effect to strengthen
the SPP, especially when QPT is relevant. Now that the enhancement of the
SPP coupling is approaching QPT regime\cite{Wallraff2004, Gunter2009,
Niemczyk2010, Peropadre2010, Forn-Diaz2010, Cristofolini2012, Scalari2012,
Xiang2013,Yoshihara2017,Andersen2017}, the TPP-SPP competition calls for a
new light in the context of QPT.

In this work we try to elucidate the role of TPP in QPT by extracting the
full ground-state phase diagram of a general model with mixed SPP and TPP.
In the weak SPP coupling regime, indeed the TPP does not come to effect
unless the TPP coupling is very strong. It would seem natural to speculate
that strengthening the SPP coupling further should relatively dwarf the TPP
more. However, we find counter-intuitively that in the strong SPP regime
properties may become very sensitive to the TPP in the sense that even a
tiny TPP coupling may bring about a dramatic change, appearing as a
spontaneous symmetry breaking. A two-face scenario emerges with both
continuous and discontinuous features for the QPT, which is found to have
different nature from two hidden transitions. Finite frequencies separates
and show the true colors of the two transitions, to which different physical
properties are sensitive respectively. We clarify the subtle mechanisms,
thus unveiling the underlying role of TPP and its entanglement with the SPP.

\textit{The model.--}The QRM with both SPP and TPP can be implemented in
superconducting circuits\cite%
{Pedernales-PRL-2018,Bertet-Nonlinear-Experim-Model-2005}. We consider
Hamiltonian \cite{biasH}
\[
H=\omega a^{\dag }a+\frac{\Omega }{2}\sigma _{x}+g_{1}\sigma _{z}{(a^{\dag
}+a)}+g_{2}\sigma _{z}{\left[ (a^{\dag })^{2}+a^{2}+\chi \widetilde{n}\right]
,}
\]%
where $\sigma _{x,y,z}$ is the Pauli matrix, $a^{\dagger }(a)$ creates
(annihilates) a bosonic mode with frequency $\omega $ and $g_{1}$,$g_{2}$
are the coupling strengths in the SPP and TPP. Here we have introduced a
Stark-like term\cite{Felicetti2018-mixed-TPP-SPP,Peter-2017} with\ $%
\widetilde{n}=a^{\dag }a+aa^{\dag }$ to retrieve the conventional TPP case
\cite{Pedernales-PRL-2018,Felicetti2018-mixed-TPP-SPP} by $\chi =0$ and\ the
quadratic form ${(a^{\dag }+a)}^{2}$ in experimental setups \cite%
{Bertet-Nonlinear-Experim-Model-2005} by $\chi =1$. We have adopted the spin
notation before the basis transformation \cite{Felicetti2018-mixed-TPP-SPP},
thus the $\Omega $ term effectively plays a role of tunneling between the
spin-up and spin-down states in $z$ direction \cite{Ying2015,Irish2014}.

\begin{figure}[t]
\includegraphics[width=1.0\columnwidth]{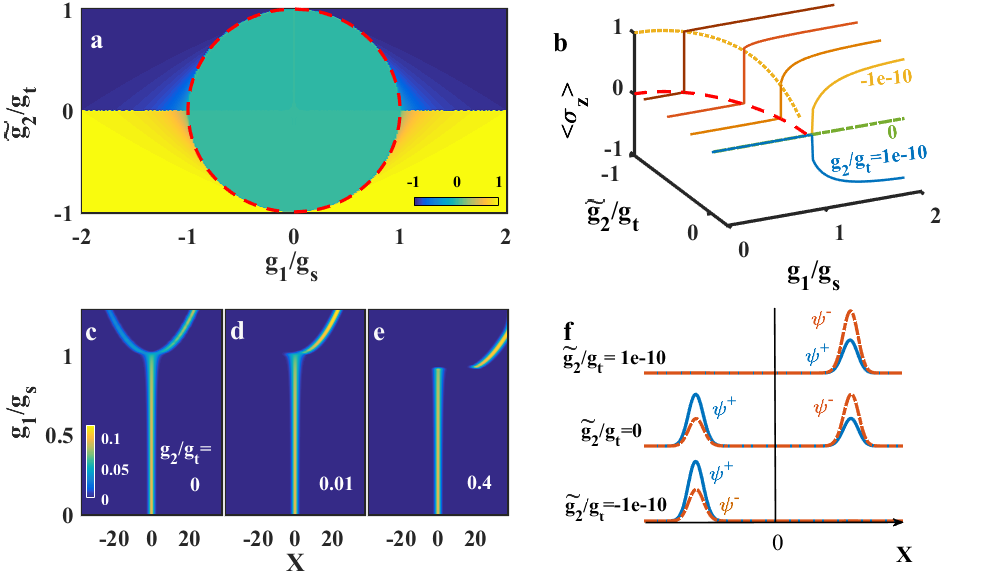}
\caption{(color online) (a) Phase diagram of $\left\langle \protect\sigma %
_{z}\right\rangle $ in $g_{1}$-$g_{2}$ plane at $\protect\omega /\Omega
=0.001$ and $\widetilde{g}_{2}=(1+\protect\chi )g_{2}$. The dashed line
represents our analytic $g_{1c}$. (b) 3-dimensional (3D) view for $%
\left\langle \protect\sigma _{z}\right\rangle $ with $\left\langle \protect%
\sigma _{z}\right\rangle =0$ at $g_{2}=0$ while there is a jump around $%
g_{2}=0$ for $g_{1}>g_{\mathrm{s}}$. (c-e) Evolution of the spin-down
function $\protect\psi ^{-}$ versus $g_{1}$ at $g_{2}/g_{\mathrm{T}%
}=0,0.01,0.3$. (f) Profile of $\protect\psi ^{+}$ and $\protect\psi ^{-}$ at
$g_{2}/g_{\mathrm{t}}=0,\pm 10^{-10}$ and $g_{1}=1.5g_{\mathrm{s}}$.}
\label{fig1-Sz}
\end{figure}

\textit{Spontaneous symmetry breaking.--}It turns out that enhancement of
the SPP might not suppress the role of TPP. Let us examine the spin
expectation $\left\langle \sigma _{z}\right\rangle $ starting in low
frequency limit where QPT is involved in the QRM\cite%
{Ashhab2010,Ying2015,Hwang2015PRL,Ref-gc0}, as illustrated by $\omega
/\Omega =0.001$ in Fig.\ref{fig1-Sz}. In the absence of TPP, $\left\langle
\sigma _{z}\right\rangle $ remains vanishing at any $g_{1}$ as shown by the
zero line at $g_{2}=0$ in Fig.\ref{fig1-Sz}(b), due to the parity symmetry ($%
[P,H]=0$ at $g_{2}=0$ for parity $P=e^{i\pi a^{\dagger }a/2}\sigma _{x}$).
At weak SPP couplings this vanishing-$\left\langle \sigma _{z}\right\rangle $
feature is unaffected by the TPP, however in strong-SPP regime even a tiny
value of $g_{2}$ may destroy this picture. In fact, as shown by $g_{2}\sim
10^{-10}g_{\mathrm{t}}$ where $g_{\mathrm{t}}=\omega /2,$ beyond a critical
point $g_{\mathrm{s}}=\sqrt{\omega \Omega }/2$ \cite{Ashhab2010,Ref-gc0} $%
\left\langle \sigma _{z}\right\rangle $ jumps away from the zero line,
either positively or negatively depending on the sign of $g_{2}$. Note that,
instead of a linear response to $g_{2}$, this jump is abrupt, which is a
characteristic behavior of \textit{spontaneous} symmetry breaking. This
contrast indicates that increasing the SPP coupling may make the parity
symmetry more vulnerable to the TPP, contrary to the intuitive speculation
that the TPP effect should be relatively weakened. As later discussed, this
peculiar trend involves a subtle TPP-SPP entanglement.

Monitoring the evolution of the wave function will make us more recognize
the essential change in the nature of quantum state. Indeed, as shown in Fig.%
\ref{fig1-Sz}(c), in the absence of TPP, below (above) $g_{\mathrm{s}}$ the
system is in a \textit{single-branch} state (\textit{double-branch} state),
which has a single wave packet (two separated wave packets) in the wave
function $\left\vert \psi \right\rangle =\psi ^{+}\left\vert \uparrow
\right\rangle -\psi ^{-}\left\vert \downarrow \right\rangle $ with spin-$z$
components $\psi ^{\pm }.$ At $g_{2}=0$ the profile of $\psi ^{\pm }\left(
x\right) $ is symmetric under simultaneous sign exchanges of effective
displacement $x=({a^{\dag }+a)}/\sqrt{2}$ and spin, as in Fig.\ref{fig1-Sz}%
(f), leading to vanishing $\left\langle \sigma _{z}\right\rangle $. In the
presence of TPP, the situation is similar before the transition, but the
transition leads to a \textit{broken-branch} state in which only one branch
survives for both $\psi ^{\pm }$, as illustrated in Fig.\ref{fig1-Sz}(d-f).
Note here that broken-branch state occurs even at a tiny TPP strength as$\
g_{2}\sim 10^{-10}g_{\mathrm{t}}$.

We extract the phase diagram of $\left\langle \sigma _{z}\right\rangle $ by
exact diagonalization in Fig.\ref{fig1-Sz}(a). We find that $\left\langle
\sigma _{z}\right\rangle $ is vanishingly small until the border \cite%
{Supplimentary} $\left\vert g_{1c}\right\vert =g_{\mathrm{s}}\sqrt{1-%
\widetilde{g}_{2}^{2}/g_{\mathrm{t}}^{2}}$ with $\widetilde{g}_{2}=(1+\chi
)g_{2}.$ We see that different $\chi $ cases can be scaled into a unified
phase diagram. Beyond the border $\left\langle \sigma _{z}\right\rangle $
jumps to finite values, and there is a sign change across the linear QRM
line at $g_{2}=0,$ while along this line $\left\langle \sigma
_{z}\right\rangle $ remains zero which is discontinuous from the regions
beside as afore-discussed. Such spontaneous symmetry breaking behavior can
also be seen in other physical quantities, e.g. the displacement expectation
$\langle a^{\dag }+a\rangle $.\cite{Supplimentary}
\begin{figure}[t]
\includegraphics[width=1.0\columnwidth]{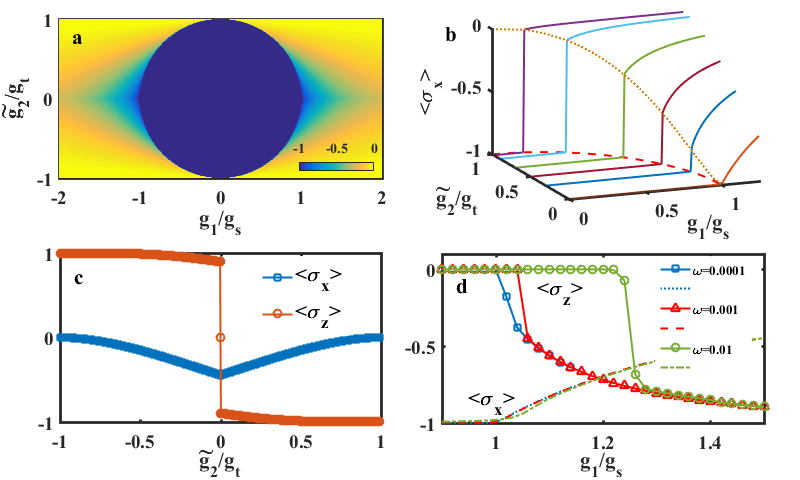}
\caption{(color online) (a,b) Phase diagram and 3D view of $\left\langle
\protect\sigma _{x}\right\rangle $ at $\protect\omega /\Omega =0.001$. (c) $%
\left\langle \protect\sigma _{z}\right\rangle $ and $\left\langle \protect%
\sigma _{x}\right\rangle $ versus $g_{2}$ at fixed $g_{1}=1.5g_{\mathrm{s}}$%
. (d) $\left\langle \protect\sigma _{z}\right\rangle $ (symbols) and $%
\left\langle \protect\sigma _{x}\right\rangle $ (lines) versus $g_{1}$ with $%
\protect\omega =0.0001,0.001,0.01$ at $g_{2}=10^{-13}g_{\mathrm{s}}$. }
\label{fig2-Sx}
\end{figure}

\textit{Continuity-discontinuity double faces in transition.--}The spin
expectation $\left\langle \sigma _{x}\right\rangle $ yields another
scenario. Figures \ref{fig2-Sx}(a,b) illustrate the behavior of $%
\left\langle \sigma _{x}\right\rangle $ for $\omega /\Omega =0.001.$ At $%
g_{2}=0$, the weak coupling regime remains flat as $\left\langle \sigma
_{x}\right\rangle \sim -1$, while $\left\langle \sigma _{x}\right\rangle $
starts to rise once beyond $g_{1c}$. The continuous change of $\left\langle
\sigma _{x}\right\rangle $ at $g_{1c}$ indicates a phase transition of
second order. At finite $g_{2}$, the transition becomes discontinuous (first
order) with a jump of $\left\langle \sigma _{x}\right\rangle $ at $g_{1c}$.
Let us denote this jump by $\Delta _{c}^{\sigma _{x}}$ while the counterpart
in $\left\langle \sigma _{z}\right\rangle $ by $\Delta _{c}^{\sigma _{x}}.$
Although phase diagram of $\left\langle \sigma _{x}\right\rangle $ seems to
show a \textquotedblleft same\textquotedblright\ transition boundary $g_{1c}$
as $\left\langle \sigma _{z}\right\rangle $, a detailed comparison reveals\
characters different from $\left\langle \sigma _{z}\right\rangle $.
On the one hand, by switching on $g_{2}$, for $\left\langle \sigma
_{x}\right\rangle $ the transition at $g_{1c}$\ evolves continuously from
second order to first order so that $\Delta _{c}^{\sigma _{x}}$ increases
smoothly ($\Delta _{c}^{\sigma _{x}}=2\widetilde{g}_{2}^{2}/(\widetilde{g}%
_{2}^{2}+g_{\mathrm{t}}^{2})$ \cite{Supplimentary} as in Fig.\ref{fig2-Sx}%
(b)), in a contrast to $\Delta _{c}^{\sigma _{z}}$ which follows $\Delta
_{c}^{\sigma _{z}}=-2\widetilde{g}_{2}g_{\mathrm{t}}/(\widetilde{g}%
_{2}^{2}+g_{\mathrm{t}}^{2})$ at finite $g_{2}$ but shows a jump upon
turning on $g_{2}$. On the other hand, beyond $g_{1c}$ the dependence of $%
\left\langle \sigma _{x}\right\rangle $ on $g_{2}$ is continuous, only
discontinuous in the derivative with respect to $g_{1}$ due to the cusp
around $g_{2}=0$ (see Fig.\ref{fig2-Sx}(c)), unlike $\left\langle \sigma
_{z}\right\rangle $ which itself is already discontinuous.

We find different nature for the double faces of discontinuity and
continuity in $\left\langle \sigma _{z}\right\rangle $ and $\left\langle
\sigma _{x}\right\rangle $. Actually rather than one there are two
transitions to which $\left\langle \sigma _{z}\right\rangle $ and $%
\left\langle \sigma _{x}\right\rangle $ are sensitive respectively. These
two transitions are so close at very low frequencies that they seem to be
born at the same $g_{1c}$, which accounts for the
double-faced behavior. Nevertheless, the two transitions will be detached
from each other at finite frequencies, as indicated in Fig.\ref{fig2-Sx}(d)
where the transition point of $\left\langle \sigma _{z}\right\rangle $ is
moving under frequency variation while that of $\left\langle \sigma
_{x}\right\rangle $ is unaffected.
\begin{figure}[t]
\includegraphics[width=1.0\columnwidth]{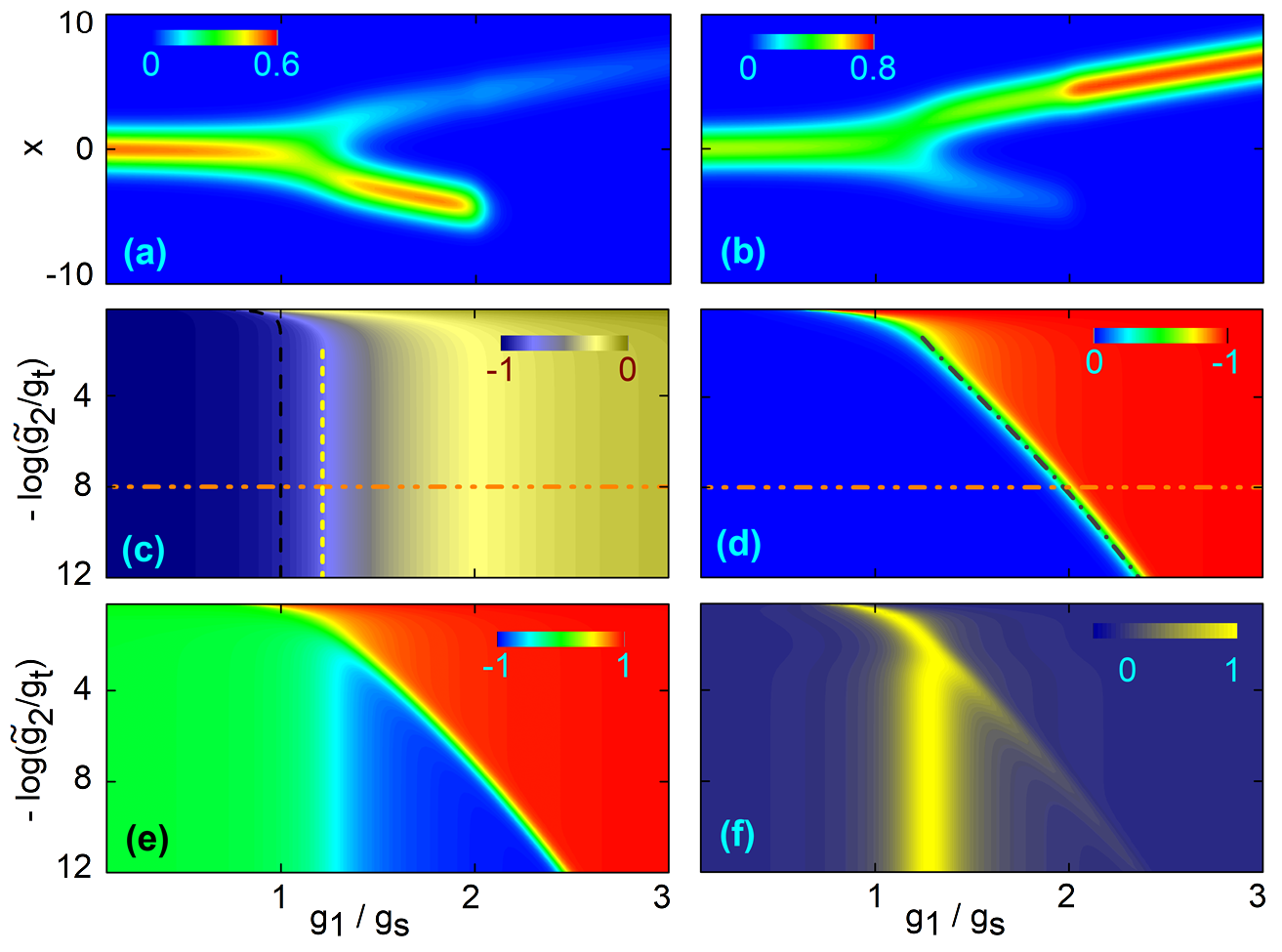}
\caption{(color online). (a,b) Evolution of $\protect\psi ^{+}$ and $\protect%
\psi ^{-}$ versus $g_{1}$ at $\protect\omega /\Omega =0.1$ and $%
g_{2}=10^{-8}g_{\mathrm{t}}$ (horizontal line in (c,d)). (c-f) Phase
diagrams of $\left\langle \protect\sigma _{x}\right\rangle $, $\left\langle
\protect\sigma _{z}\right\rangle $, $\widetilde{x}_{+}$ and $d{\widetilde{x}%
_{-}}/dg_{1}$. $d{\widetilde{x}_{-}}/dg_{1}$ is scaled by its maximum along $%
g_{1}$. Here $g_{1},g_{2}>0$, other quadrants are either symmetric or
antisymmetric\protect\cite{Supplimentary}. The dashed-, short-dashed- and
dot-dashed lines in (c,d) are our analytic $g_{1c}$ ($\protect\omega %
\rightarrow 0$), $g_{1c}^{\mathrm{I}}$ and $g_{2c}^{\mathrm{II}}$ (finite $%
\protect\omega $). }
\label{fig3-gC1gC2-w01}
\end{figure}

\textit{Two successive transitions at finite frequencies.--}To clearly show
the true colors of the transitions of $\left\langle \sigma _{z}\right\rangle
$ and $\left\langle \sigma _{x}\right\rangle $, we tune up the frequency and
give an illustration at $\omega /\Omega =0.1$ where the two transitions part
company enough. Figures \ref{fig3-gC1gC2-w01}(a,b) show the wave-function
evolution in increasing $g_{1}$ at a fixed $\widetilde{g}_{2}=10^{-8}g_{%
\mathrm{t}}$. It can be clearly seen that there are three distinct regimes,
respectively in single-, double- and broken-branch states. We denote the
boundary between single- and double-branch regimes by $\{g_{1c}^{\mathrm{I}},%
\widetilde{g}_{2c}^{\mathrm{I}}\}$ and that between double- and
broken-branch regimes by $\{g_{1c}^{\mathrm{II}},\widetilde{g}_{2c}^{\mathrm{%
II}}\}$. Figures \ref{fig3-gC1gC2-w01}(c,d) demonstrate different responses
of $\left\langle \sigma _{x}\right\rangle $ and $\left\langle \sigma
_{z}\right\rangle $ where the phase diagrams show two boundaries,
respectively, $g_{1c}^{\mathrm{I}}$ in $\left\langle \sigma
_{x}\right\rangle $ while $g_{1c}^{\mathrm{II}}$ in $\left\langle \sigma
_{z}\right\rangle $. Beyond $g_{1c}^{\mathrm{I}}$ a quick increase is
visualized in $\left\langle \sigma _{x}\right\rangle $ while no obvious
change is seen at $g_{1c}^{\mathrm{I}}$; Reversely, $\left\langle \sigma
_{z}\right\rangle $ remains vanishingly small and untouched until it jumps
to a finite value at $g_{1c}^{\mathrm{II}},$ without leaving any imprint at $%
g_{1c}^{\mathrm{I}}$. This indicates the transitions of $\left\langle \sigma
_{x}\right\rangle $ and $\left\langle \sigma _{z}\right\rangle $ have
different nature indeed. There are other observables that are respectively
sensitive to the two transitions. For examples, the mean photon number $%
\left\langle a^{\dagger }a\right\rangle ,$ the SPP coupling correlation $%
\langle \sigma _{z}({a^{\dag }+a)\rangle },$ the squeezing ratio $\langle
\widehat{p}^{2}\rangle /\langle \widehat{p}^{2}\rangle _{0}$ \cite{ashhab2},
are all among the group of $\left\langle \sigma _{x}\right\rangle $ that
exhibits obvious changes around $g_{1c}^{\mathrm{I}}$, while the
displacement expectation $\langle {a^{\dag }+a\rangle }$ and the TPP
coupling correlation $\langle \sigma _{z}{[(a^{\dag })^{2}+a^{2}]\rangle }$
\ join the group of $\left\langle \sigma _{z}\right\rangle $ that manifests
a transition behavior at $\widetilde{g}_{1c}^{\mathrm{II}}$.

It is also possible to detect both transitions by one physical quantity. We
propose the renormalized spin-filtered displacement $\widetilde{x}_{\pm
}=\langle {a^{\dag }+a\rangle }_{\pm }/(\sqrt{2}\rho _{\pm }\left\vert
x_{0,sign(-\widetilde{g}_{2})}\right\vert )$, where $\rho _{\pm }=\langle
\psi ^{\pm }|\psi ^{\pm }\rangle $ is the spin-component weight which is
related to spin expectation via $\rho _{\pm }=\left( 1\pm \left\langle
\sigma _{z}\right\rangle \right) /2$ and $x_{0,\pm }=\mp g_{1}^{\prime
}/(1\pm \widetilde{g}_{2}^{\prime }),$ with $g_{1}^{\prime }=\sqrt{2}%
g_{1}/\omega \ $\ and $\widetilde{g}_{2}^{\prime }=2\widetilde{g}_{2}/\omega
,$ is the potential displacement. The color contrast of the $\widetilde{x}%
_{+}$ map in Fig.\ref{fig3-gC1gC2-w01}(e) clearly shows three regions with
the two boundaries below a \textit{triple point}. Indeed, by increasing $%
g_{1}$ at a fixed $\widetilde{g}_{2}$ the value of $\widetilde{x}_{+}$ \ is
small and varies little in the first region, but starts to increase fast
after entering the second region, and transits to the third region with a
sign reversion. $\widetilde{x}_{-}$ does not change sign in this $g_{1}$-$%
\widetilde{g}_{2}$ quadrant but shows some clue in gradient around both
boundaries, as indicated by the local peaks of $d\widetilde{x}_{-}/dg_{1}$
in Fig.\ref{fig3-gC1gC2-w01}(f).

\begin{figure}[tbp]
\includegraphics[width=1.0\columnwidth]{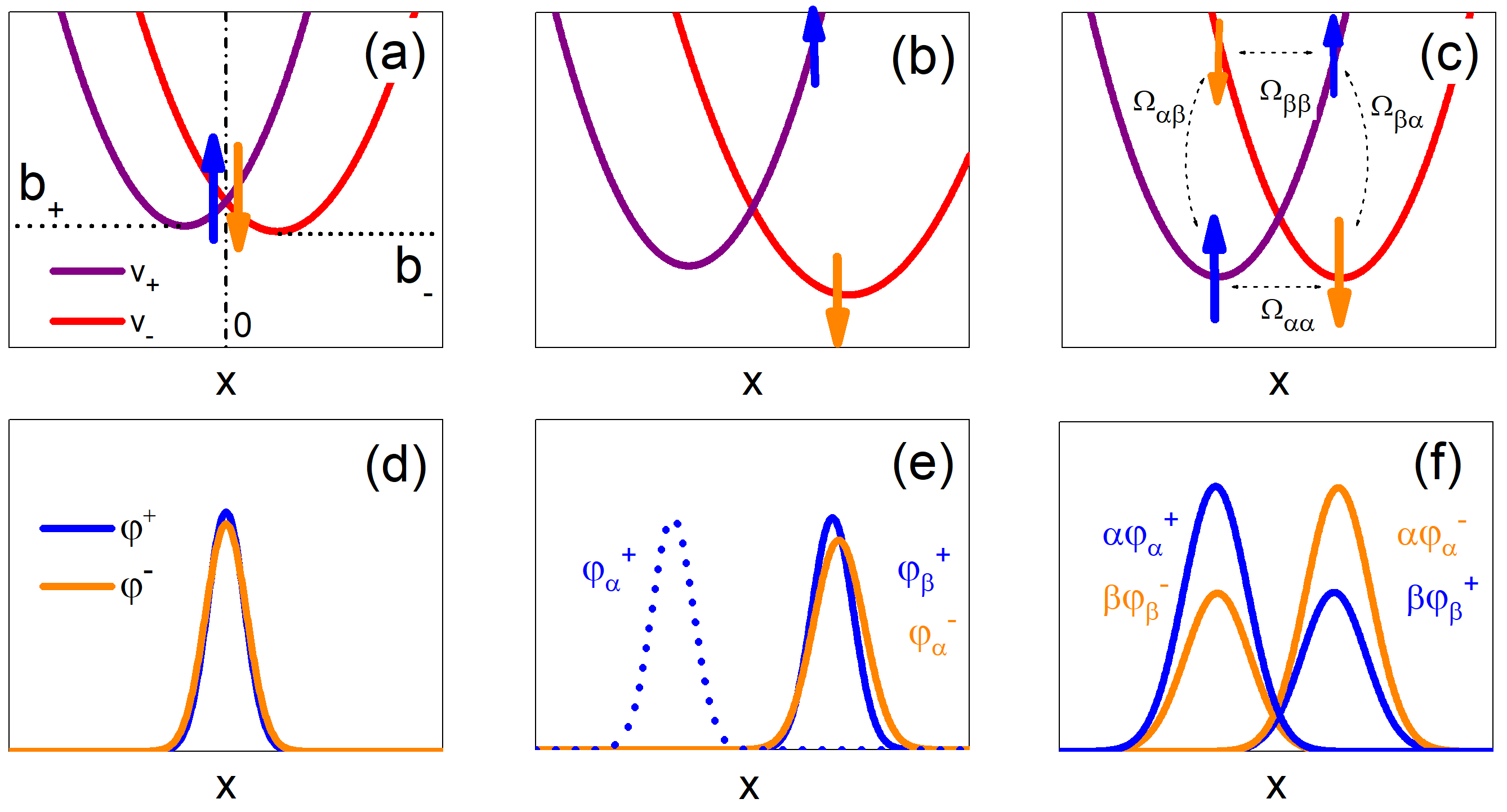}
\caption{(color online) Potential $v_{\pm }$ (upper graphs) and wave packets
(lower graphs) for single-(a,d), broken-(b,e) and double-branch (c,f)
states. The arrows mark the spin locations in $v_{\pm }$. To facilitate
understanding, as in (c,f), one can decompose the wave packet into lower- ($%
\protect\varphi _{\protect\alpha }$) and higher-potential ($\protect\varphi %
_{\protect\beta }$) parts, $\protect\psi ^{\pm }=\protect\alpha ^{\pm }%
\protect\varphi _{\protect\alpha }^{\pm }+\protect\beta ^{\pm }\protect%
\varphi _{\protect\beta }^{\pm }$ with weights $\protect\alpha ^{\pm },%
\protect\beta ^{\pm }$ for $\protect\varphi _{\protect\alpha ,\protect\beta %
}^{\pm }$ approximated by displaced harmonic oscillator\protect\cite%
{Ying2015,Supplimentary}. Thus there are four channels of tunneling energies
$\Omega _{ij}=w_{ij}\Omega S_{i\overline{j}}/2$ with $i,j\in \{\protect%
\alpha ,\protect\beta \}$ and $w_{ij}$ is product of $\protect\alpha ^{\pm },%
\protect\beta ^{\pm }$. $S_{i\overline{j}}=\langle \protect\varphi _{i}^{+}|%
\protect\varphi _{j}^{-}\rangle $ is the wavepacket overlap as sketched in
(e). Dotted line in (e) shows the left-right overlap vanishing in the
broken-branch state, while it is finite in (f). }
\label{Fig4-mechanisms} 
\end{figure}

\textit{Underlying mechanisms.--}It will facilitate the understanding if we
rewrite Hamiltonian $H$ in terms of the quantum harmonic oscillator $%
a^{\dagger }=(\hat{x}-i\hat{p})/\sqrt{2}$, $a=(\hat{x}+i\hat{p})/\sqrt{2}$,
where $\hat{x}$ and $\hat{p}$ are position and momentum, as $H=\sum_{\sigma
_{z}=\pm }(h^{\sigma _{z}}|\sigma _{z}\rangle \langle \sigma _{z}|+{\frac{%
\Omega }{2}}|\sigma _{z}\rangle \langle \overline{\sigma }_{z}|),$ where $%
\overline{\sigma }_{z}=-\sigma _{z}$ and $+$($-$) labels the up $\uparrow $
(down $\downarrow $) spin,%
\begin{equation}
h^{\pm }=\omega \ (\hat{p}^{2}/2m_{\pm }+v_{\pm })+\varepsilon _{0},\quad
v_{\pm }=v_{\pm }^{\mathrm{hp}}+b_{\pm },
\end{equation}%
where $m_{\pm }=\left( 1\mp g_{2}^{\prime }\pm \chi g_{2}^{\prime }\right)
^{-1}$ is effective mass and $\varepsilon _{0}=-[g_{1}^{\prime 2}/(1-%
\widetilde{g}_{2}^{\prime 2})+1]\omega /2$ is a constant$.$ Although the
harmonic potential $v_{\pm }^{\mathrm{hp}}=m_{\pm }\varpi _{\pm
}^{2}[x-x_{0,\pm }]^{2}/2$ has a frequency $\varpi _{\pm }=[\left( 1\pm \chi
g_{2}^{\prime }\right) ^{2}-g_{2}^{\prime 2}]^{1/2}$ degenerate at $\chi =0$%
, the TPP leads to an effective bias field $b_{\pm }=\pm \widetilde{g}%
_{2}^{\prime }g_{1}^{\prime 2}/[2(1-\widetilde{g}_{2}^{\prime 2})]$ which
tends to raise the degeneracy at any value of $\chi $. We see that, although
induced by the TPP, the bias $b_{\pm }$ becomes larger if one increases the
SPP strength $g_{1}$. This entangled bias leads to different transition
scenarios above and below the triple point.

In reality, whether degeneracy raising comes into final effect depends on
what quantum state the system is in. As indicated by $x_{0,\pm }$ in $v_{\pm
}^{\mathrm{hp}}$, one effect of the linear interaction $g_{1}$ is to
separate the potentials. However, the relative large tunneling strength $%
\Omega /\omega $ at low frequencies prevents the true separation of spin-up
and spin-down wave packets $\psi ^{+}$ and $\psi ^{-}$, as depicted in Fig.%
\ref{Fig4-mechanisms}(a), since larger wave-packet overlap gains more
negative tunneling energy. As a result, the two spin components will stay
together at the origin $x=0$ to form a single-branch state before
transition. It happens that the harmonic potential at the origin, $v_{\pm }^{%
\mathrm{hp}}\left( 0\right) =(1\mp \widetilde{g}_{2}^{\prime })g_{1}^{\prime
2}/[2(1-\widetilde{g}_{2}^{\prime 2})],$ cancels with the bias energy $%
b_{\pm }$ so that $\psi ^{+}$ and $\psi ^{-}$ have a same potential. The
ground state in low-frequency limit can be described by semiclassical
consideration\cite{Ashhab2010,Supplimentary} with $\hat{p}^{2}\rightarrow 0$%
, which yields a degenerate energy, accounting for the vanishingly-small $%
\left\langle \sigma _{z}\right\rangle $ in single-branch state. Note $m_{\pm
}\varpi _{\pm }^{2}=\left( 1\pm \widetilde{g}_{2}^{\prime }\right) $ in $%
v_{\pm }^{\mathrm{hp}},$\ the energy then is a function of $\widetilde{g}%
_{2}^{\prime }$, thus leading to the afore-mentioned scaling behavior of $%
\chi $. Further enhancement of the SPP with larger $g_{1}$ will also enlarge
the bias gap, as depicted in Fig.\ref{Fig4-mechanisms}(b), due to the
entanglement of $g_{1}$ and $g_{2}$. Once the tunneling energy gain at the
origin cannot afford the high potential cost in more separated potential $%
v_{\pm }^{\mathrm{hp}}$, a transition occurs from single-branch state to
broken-branch state at $g_{1c}$.

In the case of a weak TPP below the triple point, the energy competitions
are more separated. The bias opening is slowed down by small $g_{2}$ so that
the potential cost resulting from $g_{1}$-driven horizontal separation of $%
v_{\pm }^{\mathrm{hp}}$ dominates first in competition with the tunneling,
which leads to transition to double-branch state around $g_{1c}^{\mathrm{I}%
}\approx \sqrt{\omega ^{2}+\sqrt{\omega ^{4}+g_{\mathrm{s}}^{4}}}$\cite%
{Supplimentary,Ying2015}. In this situation, as in Fig.\ref{Fig4-mechanisms}%
(c), even the weak left-right tunneling $\Omega _{\alpha \alpha },\Omega
_{\beta \beta }$ can balance the two-side distribution due to the small bias.

Increasing $g_{1}$ deeper eventually becomes detrimental to the balance: $%
g_{1}$ not only separates $v_{\pm }^{\mathrm{hp}}$ more thus leading to a
faster decay in left-right\ tunneling, but also is enlarging the bias. This
triggers the second transition between double- and broken-branch\ states.
Around this transition, the final state $\left\vert \Psi \right\rangle =$ $%
\left\vert \psi _{\mathrm{R}}\right\rangle +\delta _{c}\left\vert \psi _{%
\mathrm{L}}\right\rangle $ is a superposition of the right/left states $\psi
_{\mathrm{R,L}}$, with energy $\varepsilon _{\mathrm{R,L}}$ in same-side
tunneling $\Omega _{\alpha \beta }$ and $\Omega _{\alpha \beta }$, by a
perturbation $\delta _{c}=(\Omega _{\alpha \alpha }+\Omega _{\beta \beta
}+t_{\alpha \beta }^{+}+t_{\beta \alpha }^{-})/\left( \varepsilon _{\mathrm{L%
}}-\varepsilon _{\mathrm{R}}\right) $ from left-right tunneling $\Omega
_{ii} $ as well as a smaller contribution from single-particle off-diagonal
energy $t_{ij}^{\pm }$ \cite{footnotes}, which leads us to the analytic
second boundary \cite{Supplimentary} at small $g_{2}$ and finite
frequencies, traced by an exponantial weight decay $\delta _{c}\sim e^{-1}$:%
\begin{equation}
\left\vert \overline{g}_{2c}^{\mathrm{II}}(g_{1})\right\vert \approx \frac{%
\exp [-\zeta ^{2}\overline{g}_{1}^{2}\Omega /(2\omega )]}{\delta _{c}\zeta
^{3}\overline{g}_{1}^{2}(1-t)^{-1}},\ \overline{g}_{1}\equiv \frac{g_{1}}{g_{%
\mathrm{s}}},\ \overline{g}_{2}\equiv \frac{\widetilde{g}_{2}}{g_{\mathrm{t}}%
},
\end{equation}%
where $t=(1-\zeta )^{2}/2+\omega /(\overline{g}_{1}^{2}\Omega )$ is the $%
t_{ij}^{\pm }$ contribution and $\zeta =(1-\overline{g}_{1}^{-4})^{1/2}$ is
displacement renormalization \cite{Ying2015}. $\overline{g}_{2c}^{\mathrm{II}%
}$ is plotted as the dot-dashed line in Fig.\ref{fig3-gC1gC2-w01}(d). Given
a SPP coupling $\overline{g}_{1}$ the TPP transition point $\overline{g}%
_{2c}^{\mathrm{II}}$ can be tuned to a measurable order by raising the
frequency \cite{Supplimentary}. We see that $\overline{g}_{2c}^{\mathrm{II}}$
is fully a quantum-mechanics effect, the left-right wave-packet overlap
cannot be captured by a semiclassical consideration with a mass point.%

Note that in the broken-branch state $\left\vert \psi _{\mathrm{R}%
}\right\rangle $, the spin-up component has less weight than spin-down due
to potential\ imbalance, thus leading to finite $\left\langle \sigma
_{z}\right\rangle ,$ while in the double-branch state this spin imbalance
cancels between $\left\vert \psi _{\mathrm{R}}\right\rangle $ and $%
\left\vert \psi _{\mathrm{L}}\right\rangle $ thus remaining in a vanishing $%
\left\langle \sigma _{z}\right\rangle $ as in the single-branch state. This
is the reason why $\left\langle \sigma _{z}\right\rangle $ is sensitive to
the second transition but responseless to the first one. As for $%
\left\langle \sigma _{x}\right\rangle $, the potential imbalance is opened
in the transition from single-branched state to the double-branch state,
which leads to a quick reduction of $\psi ^{+}$-$\psi ^{-}$overlap, giving a
fast decrease of the spin flipping i.e. the $\left\langle \sigma
_{x}\right\rangle $ amplitude at the first transition. But in shifting from
double-branch state to broken-branch state the sum of leading same-side
overlaps $S_{\alpha \beta }^{+-}$ and $S_{\alpha \beta }^{+-}$ is not
affected, thus no sign of the second transition is observed in $\left\langle
\sigma _{x}\right\rangle $. Concerning the low frequency limit, the much
narrower wave packets relative to the packet distance result in an immediate
decay of left-right overlap in the $g_{1}$-driven wave packet splitting,
thus the two transitions become less separated.

Conclusions.\textit{--}We have seen that the TPP-SPP interplay leads to an
entangled effective bias, which tends to raise spin degeneracy without an
external field and brings about a first-order transition, in \ a way that
strengthening SPP does not dwarf but enhances the role of TPP. In a strong
SPP even a tiny strength of TPP can be crucial and lead to a spontaneous
symmetry breaking behavior, this scenario could survive in the presence of
external/environmental modes despite the intuition that the tiny TPP
parameter seemingly should be negligible in comparison with the finite
couplings to the external modes\cite{Supplimentary}.
At finite frequencies we unveil two successive transitions hidden in the
weak TPP regime, with the first-order transition in strong TPP splitting
into second-order- and first-order-like ones below a triple point. Different
groups of physical quantities are distinguished\ to be sensitive,
respectively or simultaneously, to these transitions, thus useful for
detections. The clarified mechanism shows a delicate competition of the TPP,
SPP and tunneling. Note that superconducting qubits can be easily cooled to
the ground state\cite{Forn-Diaz2010}, our results are relevant for enhanced
SPP couplings in rapid experimental progress\cite{Wallraff2004, Gunter2009,
Niemczyk2010,FornDiaz2017, Peropadre2010, Forn-Diaz2010, Cristofolini2012,
Scalari2012, Xiang2013,Yoshihara2017,Andersen2017} with the increasing
interest in the TPP \cite{Bertet2002-TwoPhotonProcess,
Brune1987-TwoPhotonProcess,Stufler2006-TwoPhotonProcess,Valle2010-TwoPhotonProcess,Verma2016-TwoPhotonProcess, Felicetti2015-TwoPhotonProcess,Puebla2017-TwoPhotonProcess,Felicetti2018-mixed-TPP-SPP,Pedernales-PRL-2018,Bertet-Nonlinear-Experim-Model-2005, Lange1996-SuppressSPP,Ota2011-SuppressSPP,Casanova2018npj,Puebla2019,CongLei2019,Xie2019}%
. We expect these nonlinearity phenomena might also leave imprints in
dynamics\cite{Crespi2012,Wolf2012,Hwang2015PRL} and Bloch-Siegert effect\cite%
{Forn-Diaz2010,Pietikainen2017}, which could be future works.

\textbf{Acknowledgements} Z.-J.Y. acknowledges partial financial support
from the Future and Emerging Technologies (FET) program under FET-Open Grant
No. 618083 (CNTQC). L.C. and X.-M.S. acknowledge National Science Foundation
of China (Grants No. 11325417 and No. 11674139). We thank Hong-Gang Luo for
valuable discussions.




\begin{thebibliography}{99}
\bibitem{footnotes} The left-right overlap also partially involves the
single-particle energies $t_{\alpha \beta }^{+}=\alpha ^{+}\beta
^{+}h_{\alpha \beta }^{+}$ and $t_{\beta \alpha }^{-}=\beta ^{-}\alpha
^{-}h_{\beta \alpha }^{-},$ where $h_{ij}^{\pm }=\langle \varphi _{i}^{\pm
}|(h-\varepsilon _{0})|\varphi _{j}^{\pm }\rangle ,$ which is within the
same spin component thus having subscripts different from $\Omega _{ij}$.
\end{thebibliography}
\end{document}